\documentclass[a4paper,11pt]{article}
\usepackage{pos}
\title{Momentum transfer dependence of kaon semileptonic form factor
on (10 fm)$^4$ at the physical point}
\ShortTitle{Momentum transfer dependence of kaon semileptonic form factor}

\author*[a,b]{Takeshi Yamazaki}

\author[c]{Ken-ichi~Ishikawa}

\author[b]{Naruhito~Ishizuka}

\author[b]{Yoshinobu~Kuramashi}

\author[d]{Yusuke~Namekawa}

\author[b]{Yusuke~Taniguchi\footnote{Yusuke Taniguchi passed away on July 22, 2022, while this work was in progress.}}

\author[b]{Naoya~Ukita}

\author[b]{Tomoteru~Yoshi\'e}

\affiliation[]{\normalsize{\bf \sffamily \hspace{50mm} (PACS Collaboration)}}

\affiliation[a]{Faculty of Pure and Applied Sciences, University of Tsukuba, Tsukuba, Ibaraki 305-8571, Japan}

\affiliation[b]{Center for Computational Sciences, University of Tsukuba, Tsukuba, Ibaraki 305-8577, Japan}

\affiliation[c]{Core of Research for the Energetic Universe, Graduate School of Advanced Science and Engineering, Hiroshima University, Higashi-Hiroshima, 739-8526, Japan}

\affiliation[d]{Education and Research Center for Artificial Intelligence and Data Innovation, Hiroshima University, Higashi-Hiroshima 739-8521, Japan}

\emailAdd{yamazaki@het.ph.tsukuba.ac.jp}

\abstract{
We calculate the kaon semileptonic form factors using the two 
sets of the PACS10 configuration, whose physical volumes are more than 
(10 fm)$^4$ at the physical point. The lattice spacings are 0.063 and 0.085 fm.
The configurations were generated using the Iwasaki gauge action and $N_f=2+1$ 
stout-smeared nonperturbatively $O(a)$-improved Wilson quark action. From the 
momentum transfer dependence of the form factors, we 
evaluate the slope and curvature for the form factors at the zero momentum 
transfer. Furthermore, we calculate the phase space factor, which is used
to obtain $|V_{us}|$ through the kaon semileptonic decay. These results are 
compared with previous lattice results and experimental values.
}

\FullConference{%
The 39th International Symposium on Lattice Field Theory,\\
8th-13th August, 2022,\\
Rheinische Friedrich-Wilhelms-Universität Bonn, Bonn, Germany
}

\begin{document}
\maketitle

\section{Introduction}

A violation of the unitarity of the Cabibbo-Kobayashi-Maskawa
(CKM) matrix implies the existence of physics beyond the standard model.
In the first row of the CKM matrix, 
a 2.2$\sigma$ violation is reported using
the current values for the matrix elements, $|V_{ud}|$, $|V_{us}|$, 
and $|V_{ub}|$~\cite{ParticleDataGroup:2022pth}.
To confirm whether this violation exists or not,
it is important to reduce the uncertainty of the matrix elements.

The value of $|V_{us}|$ can be determined through
the kaon semileptonic ($K_{\ell 3}$) decay process.
The $K_{\ell 3}$ decay rate $\Gamma_{K_{\ell 3}}$ is related to
$|V_{us}|$ as,
\begin{equation}
\Gamma_{K_{\ell 3}} = C_{K_{\ell 3}} ( |V_{us}| f_+(0) )^2 I^\ell_K,
\label{eq:decay_width_kl3}
\end{equation}
where 
$C_{K_{\ell 3}}$ is a known factor including the electromagnetic 
and SU(2) breaking corrections, and 
$f_+(0)$ is the $K_{\ell 3}$ form factor at zero momentum transfer $q^2 = 0$.
The phase space integral $I^\ell_K$ of the lepton $\ell$
is usually calculated from the shape of
the experimental $K_{\ell 3}$ form factors.
The value of $f_+(0)$ needs to be determined from a nonperturbative
calculation of the strong interaction, such as lattice QCD.

Precise values of $f_+(0)$ were provided by various lattice QCD 
calculations~\cite{Bazavov:2012cd,Boyle:2007qe,Boyle:2013gsa,Boyle:2015hfa,Aoki:2017spo,Bazavov:2013maa,Carrasco:2016kpy,Bazavov:2018kjg}
including ours~\cite{PACS:2019hxd}.
Recently we report updates of our calculation in Ref.~\cite{Ishikawa:2022otj},
where $f_+(0)$ in the continuum limit is obtained using
the two ensembles of the PACS10 configuration
generated on more than (10 fm)$^4$ volumes at the physical point.
Since we evaluate the form factors as a function of $q^2$,
in this report, we discuss physical quantities obtained from the $q^2$ 
dependence of the form factors, which are the slope and curvature
of the form factors at $q^2 = 0$, and also the phase space integral $I_K^\ell$
in Eq.~(\ref{eq:decay_width_kl3}).
Almost all results in this report were already presented 
in Ref.~\cite{Ishikawa:2022otj}.

\section{Result}

The Iwasaki gauge action and the nonperturbatively 
improved Wilson quark action with the six-stout link 
smearing~\cite{Morningstar:2003gk} are utilized to generate
the PACS10 configuration, which has more than (10 fm)$^4$ volume
at the physical point.
The details of the configuration generations are explained in
Refs.~\cite{Shintani:2019wai} and \cite{Ishikawa:2018jee}
for the finer and coarser lattice spacings, respectively.
The same quark action is employed in the measurement of 
the $K_{\ell 3}$ form factors.
The simulation parameters at each lattice spacing are
tabulated in Table~\ref{tab:sim_param} including 
the masses for $\pi$ and $K$, $M_\pi$ and $M_K$.

In this work we calculate the two $K_{\ell 3}$ form factors,
$f_+(q^2)$ and $f_0(q^2)$, from the $K_{\ell 3}$ 3-point functions.
Through the matrix element extracted from the 3-point functions,
$f_+(q^2)$ is defined by
\begin{eqnarray}
\langle \pi (\vec{p}_{\pi}) \left | V_{\mu} \right | K(\vec{p}_{K}) \rangle = ({p}_{K}+{p}_{\pi})_{\mu}f_{+}(q^2)+ ({p}_{K}-{p}_{\pi})_{\mu}f_{-}(q^2),
\label{eq:def_matrix_element}
\end{eqnarray}
where $q^2$ is the momentum transfer squared and $V_{\mu}$ 
is the weak vector current.
Another form factor $f_0(q^2)$ is defined by $f_+(q^2)$ and $f_-(q^2)$
as,
\begin{eqnarray}
f_{0}(q^2) = f_{+}(q^2) + \frac{-q^2}{{m^2_{K}}-{m^2_{\pi}}}f_{-}(q^2).
\label{eq:f0}
\end{eqnarray}
The calculation of the 3-point functions
is carried out using the local and conserved vector currents.
Thus, we have two form factor data obtained from the local and
conserved currents at each lattice spacing.
The local vector current is renormalized by
the renormalization factor 
$Z_V = 1/\sqrt{F^{\rm bare}_\pi(0)F^{\rm bare}_K(0)}$.
The two form factors, $F^{\rm bare}_\pi(0)$ and $F^{\rm bare}_K(0)$, are 
the unrenormalized electromagnetic form factors for $\pi$ and $K$ at $q^2 = 0$
with the local vector current.

\begin{table}[!t]
\caption{
Simulation parameters of the PACS10 configurations at the two lattice spacings.
The bare coupling ($\beta$), lattice size ($L^3\cdot T$),
physical spatial extent ($L$[fm]), lattice spacing ($a$[fm]),
the number of the configurations ($N_{\rm conf}$), and
pion and kaon masses ($m_\pi$, $m_K$) are tabulated.
  \label{tab:sim_param}
}
\begin{center}
\begin{tabular}{cccccccc}\hline\hline
$\beta$ & $L^3\cdot T$ & $L$[fm] & $a$[fm] &
$N_{\rm conf}$ & $m_\pi$[MeV] & $m_K$[MeV]\\\hline
2.00 & 160$^4$ & 10.1 & 0.063 & 20 & 138 & 505 \\
1.82 & 128$^4$ & 10.9 & 0.085 & 20 & 135 & 497 \\\hline\hline
\end{tabular}
\end{center}
\end{table}

\subsection{Form factors}

\begin{figure}[!t]
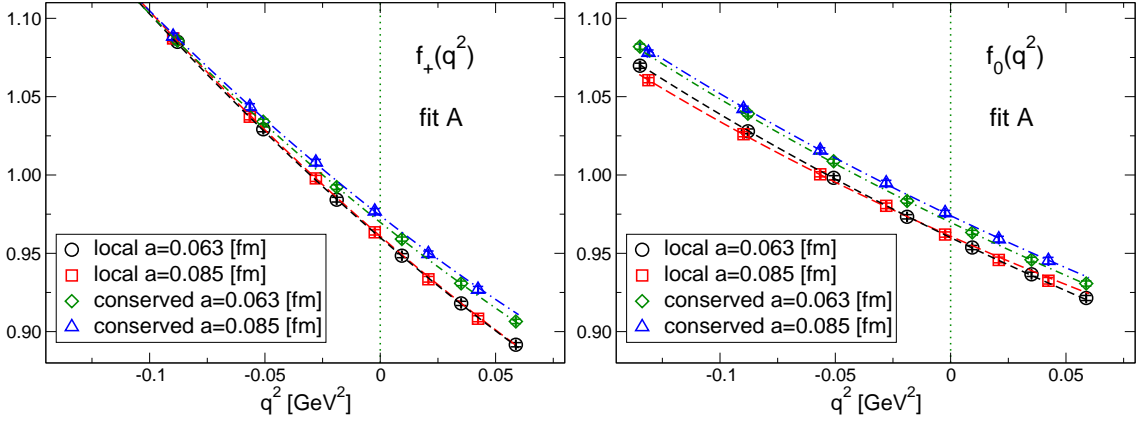

 \centering
 \includegraphics*[scale=0.43]{figs/fig-fit-fplus_q2.eps}
 \includegraphics*[scale=0.43]{figs/fig-fit-fzero_q2.eps}
 \caption{
The $K_{\ell 3}$ form factors $f_+(q^2)$ and $f_0(q^2)$ 
as a function of $q^2$ are plotted in the left and right panels,
respectively.
In each panel, the data with the local and conserved currents
at the two lattice spacings are expressed by different symbols.
The dashed and dot-dashed curves represent fit results from 
a simultaneous $q^2$ interpolation and continuum extrapolation
with fit A.
This figure is reprinted from Ref.~\cite{Ishikawa:2022otj}.
\label{fig:f+-f0-fit}
 }
\end{figure}

The data for $f_+(q^2)$ and $f_0(q^2)$ at finite lattice spacings
are presented in Fig.~\ref{fig:f+-f0-fit} as a function of $q^2$.
The figure contains the data calculated from the local and conserved
currents at the two lattice spacings.
Thanks to the huge volume of the PACS10 configurations,
we obtain several data near the $q^2 = 0$ region,
which stabilize a $q^2$ interpolation of the form factors to $q^2 = 0$.
The lattice spacing dependence of the conserved current data is larger 
than that of the local current data, especially in $f_+(q^2)$.

We carry out simultaneous $q^2$ interpolation and continuum extrapolation
using all the data we calculated: the two form factors $f_+(q^2)$ and 
$f_0(q^2)$ with the local and conserved currents at the two lattice spacings.
The fit form is based on the NLO SU(3) 
ChPT formula~\cite{Gasser:1984ux,Gasser:1984gg},
and we add correction terms corresponding to 
finite lattice spacing effects by introducing
functions $g^{\rm cur}_+(q^2,a)$ and $g^{\rm cur}_0(q^2,a)$.
The superscript ``${\rm cur}$'' expresses the local or conserved vector currents.
The fit forms are given by
\begin{eqnarray}
f_+^{\rm cur}(q^2) &=& 
1 - \frac{4}{F_0^2} L_9 q^2 + K_+(q^2) + c_0 + c_2^+ q^4
+ g^{\rm cur}_+(q^2,a),
\label{eq:ato0_NLO_chpt_f+}\\
f_0^{\rm cur}(q^2) &=& 
1 - \frac{8}{F_0^2} L_5 q^2 + K_0(q^2) + c_0 + c_2^0 q^4
+ g^{\rm cur}_0(q^2,a).
\label{eq:ato0_NLO_chpt_f0}
\end{eqnarray}
The NLO functions $K_+(q^2)$ and $K_0(q^2)$ are described in
Refs.~\cite{Gasser:1984ux,Gasser:1984gg}.
The pion decay constant in the chiral limit $F_0 = 0.11205$ GeV is employed.
We choose three types of $g^{\rm cur}_{+,0}(q^2,a)$, called fit A, B, and C, 
to investigate a systematic error of the fit results, whose explicit 
forms are described in Ref.~\cite{Ishikawa:2022otj}.

\begin{figure}[!th]
 \centering
 \includegraphics*[scale=0.43]{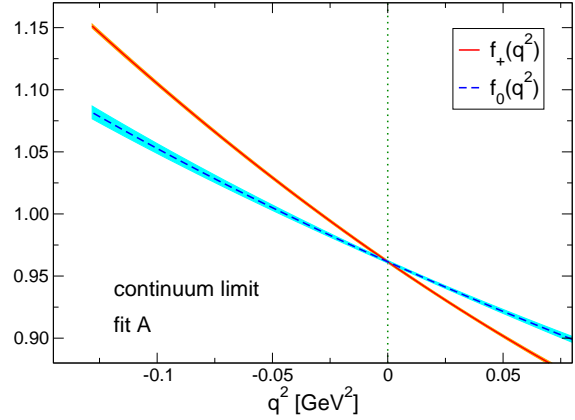}
 \caption{
The $K_{l3}$ form factors $f_+(q^2)$ and $f_0(q^2)$ in
the continuum limit using the fit A form for the continuum extrapolation.
The error bands are only statistical error.
This figure is reprinted from Ref.~\cite{Ishikawa:2022otj}.
  \label{fig:f+-f0-a0}
 }
\end{figure}

A typical example of the simultaneous fit results is presented in
Fig.~\ref{fig:f+-f0-fit} by dashed and dot-dashed curves,
which are obtained with the fit A form.
From the fit results, the form factors in the continuum limit are obtained
as a function of $q^2$ presented in Fig.~\ref{fig:f+-f0-a0}.
In this report, we will discuss physical quantities obtained from
the $q^2$ dependence of the form factors.
We note that the result of $f_+(0)$ and value of $|V_{us}|$ 
determined with our $f_+(0)$ were discussed in Ref.~\cite{Ishikawa:2022otj}.

\subsection{Slope and curvature}

\begin{figure}[!t]
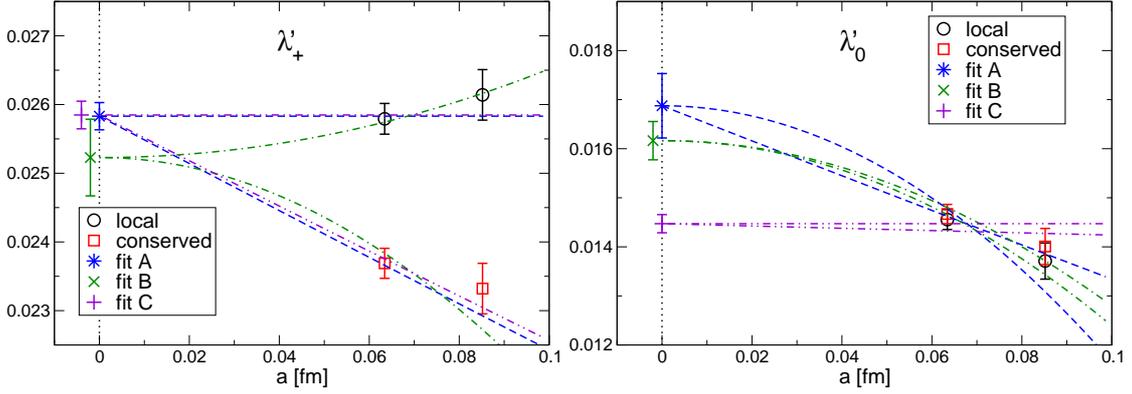

 \centering
 \includegraphics*[scale=0.41]{figs/lam+_a0.eps}
 \includegraphics*[scale=0.41]{figs/lam0_a0.eps}
 \caption{
Lattice spacing dependence of $\lambda_+^\prime$ (left)
and $\lambda_0^{\prime}$ (right).
Circle and square symbols represent the results
with the local and conserved current, respectively.
Different curves correspond to fit results using different fit forms
for the continuum extrapolation.
This figure is reprinted from Ref.~\cite{Ishikawa:2022otj}.
  \label{fig:lam+0-a0}
 }
\end{figure}

The slope and curvature of the $K_{\ell 3}$ form factors at $q^2 = 0$
are defined by
\begin{equation}
\lambda_s^\prime = \frac{m_\pi^2}{f_+(0)}\left.\frac{d f_s(q^2)}{d (-q^2)}\right|_{q^2=0} , \ \ \ 
\lambda_s^{\prime\prime} = \frac{m_\pi^4}{f_+(0)}\left.\frac{d^2 f_s(q^2)}{d (-q^2)^2}\right|_{q^2=0} ,
\end{equation}
where $s = +$ and 0.
Those quantities are evaluated from a $q^2$ fit of the form factors at each
lattice spacing and vector current data.
Figure~\ref{fig:lam+0-a0} shows the lattice spacing dependence of $\lambda_+^\prime$ and $\lambda_0^\prime$ with the two current data.
For $\lambda_+^\prime$ in the left panel, 
the two current data have a different dependence on the lattice
spacing, which is similar to the one of $f_+(0)$~\cite{Ishikawa:2022otj}.
The results in the continuum limit obtained from
the three fits using fit A, B, C, explained in the last subsection, 
are stable in this quantity.
In contrast to $\lambda_+^\prime$, 
the two current data of $\lambda_0^\prime$ in the right panel 
almost agree with each other at both lattice spacings.
Thus, the results in the continuum limit largely depend on the fit forms
in the continuum extrapolation.
A systematic error of the slopes is estimated from the differences of the 
fit results.

\begin{figure}[!t]
 \centering
 \includegraphics*[scale=0.43]{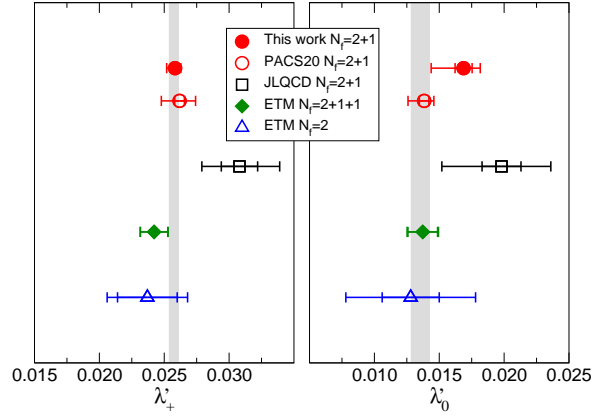}
 \caption{
Comparison of our result of $\lambda_+^\prime$ (left) and 
$\lambda_0^\prime$ (right) with previous 
lattice results~\cite{Lubicz:2009ht,Carrasco:2016kpy,Aoki:2017spo,PACS:2019hxd}
denoted by different symbols.
The closed (open) symbols represent results 
in the continuum limit (at a finite lattice spacing).
The experimental value~\cite{Moulson:2017ive} with one standard deviation
is also presented by the gray band.
The inner and outer errors express the statistical and total errors.
The total error is evaluated by adding the statistical and systematic
errors in quadrature.
This figure is reprinted from Ref.~\cite{Ishikawa:2022otj}.
  \label{fig:lam+0}
 }
\end{figure}

Our results for $\lambda_+^\prime$ and $\lambda_0^\prime$ are compared with
previous lattice calculations~\cite{Lubicz:2009ht,Carrasco:2016kpy,Aoki:2017spo,PACS:2019hxd} in Fig.~\ref{fig:lam+0}.
Our results in the continuum limit are reasonably consistent with
other lattice results and also the experimental values~\cite{Moulson:2017ive}
expressed by the gray band.
It is noted that our result of $\lambda_0^\prime$ has a larger systematic
error than that of our previous calculation using the coarser lattice spacing 
configuration denoted by the open circle,
because in the previous calculation the systematic error of 
a finite lattice spacing effect is estimated by an order counting.

\begin{figure}[!t]
 \centering
 \includegraphics*[scale=0.43]{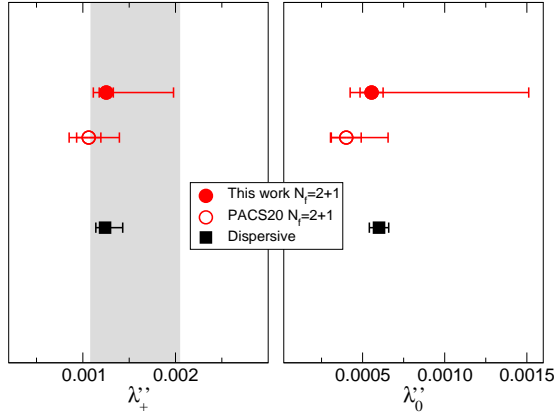}
 \caption{
Comparison of our result of $\lambda_+^{\prime\prime}$ (left) and 
$\lambda_0^{\prime\prime}$ (right) in the continuum limit 
with our previous results~\cite{PACS:2019hxd} at the coarser lattice spacing
and the experimental value evaluated with the dispersive
representation~\cite{Bernard:2009zm}.
The gray band corresponds to the average of the experimental 
results~\cite{Antonelli:2010yf}.
The inner and outer errors express the statistical and total errors.
The total error is evaluated by adding the statistical and systematic
errors in quadrature.
This figure is reprinted from Ref.~\cite{Ishikawa:2022otj}.
  \label{fig:dlam+0}
 }
\end{figure}

A similar lattice spacing dependence to that
of $\lambda_0^\prime$ is seen in both the curvatures,
so that the extrapolated results largely depend on the fit forms.
In Fig.~\ref{fig:dlam+0}, the results in the continuum limit are compared 
with our previous calculation~\cite{PACS:2019hxd}, the experimental value 
evaluated with the dispersive representation~\cite{Bernard:2009zm},
and the average of the experimental results~\cite{Antonelli:2010yf}.
Our continuum results agree with those results,
albeit our results have larger errors.

\subsection{Phase space integral}

The phase space integral $I_K^\ell$ in Eq.~(\ref{eq:decay_width_kl3}) 
is defined by an integral of 
the form factors~\cite{Leutwyler:1984je} as,
\begin{equation}
I^\ell_K = \int^{t_{\rm max}}_{m_\ell^2} dt
\frac{\lambda^{3/2}}{M_K^8}
\left( 1 + \frac{m_\ell^2}{2 t} \right)
\left( 1 - \frac{m_\ell^2}{t} \right)^2
\left(
\overline{F}_+^2(t) + 
\frac{3 m_\ell^2 \Delta_{K\pi}^2}{(2t + m_\ell^2)\lambda}\overline{F}_0^2(t)
\right),
\label{eq:phase_space_integral}
\end{equation}
where $t = -q^2$, $t_{\rm max} = ( M_K - M_\pi )^2$,
$\lambda = ( t - ( M_K + M_\pi )^2 ) ( t - t_{\rm max} )$,
$\Delta_{K\pi} = M_K^2 - M_\pi^2$,
$m_\ell$ is the mass of the lepton $\ell$, and
$\overline{F}_s(t) = f_s(-t)/f_+(0)$ with $s = +$ and 0.
We evaluate $I^\ell_{K^0}$ and $I^\ell_{K^+}$ for $\ell = e, \mu$
using ($M_K$, $M_\pi$) = ($m_{K^0}$, $m_{\pi^-}$) and ($m_{K^+}$, $m_{\pi^0}$),
respectively.

\begin{figure}[!t]
 \centering
 \includegraphics*[scale=0.43]{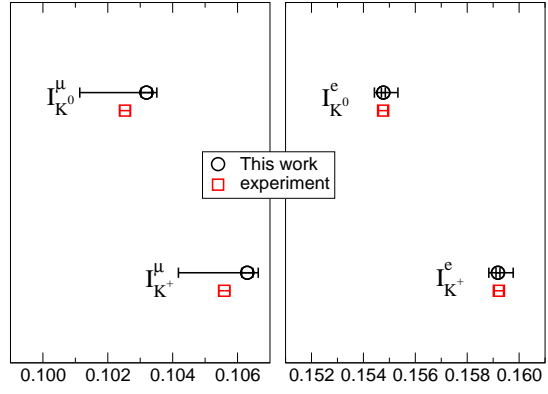}
 \caption{
Results of the phase space integral evaluated from our form factors
for the four channels.
The experimental values~\cite{Antonelli:2010yf} are also plotted.
The inner and outer errors of the circle symbols 
express the statistical and total errors.
The total error is evaluated by adding the statistical and systematic
errors in quadrature.
  \label{fig:Ikl}
 }
\end{figure}

The results of $I_K^\ell$ calculated with our form factors in
the continuum limit are shown in Fig.~\ref{fig:Ikl} for each lepton.
While our results have large systematic errors coming from
the fit form dependence in the continuum extrapolations,
those results are consistent with the experimental values 
using the form factors in the dispersive representation~\cite{Antonelli:2010yf}.

\begin{figure}[!t]
 \centering
 \includegraphics*[scale=0.43]{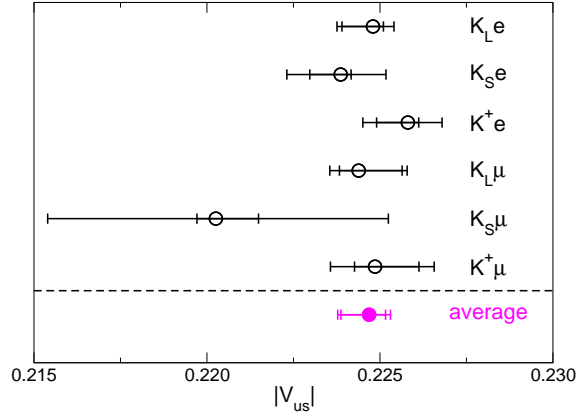}
 \caption{
Values of $|V_{us}|$ determined from our result of the phase space 
integral in the six decay processes.
The average of those values is also plotted.
The inner and outer errors
express the lattice QCD and total errors.
The total error is evaluated by adding the errors in the lattice QCD and experiment in quadrature.
  \label{fig:vus_Ikl}
 }
\end{figure}

The value of $|V_{us}|$ is determined using our result of $I_K^\ell$
for six $K_{\ell 3}$ decay processes.
The experimental value of $|V_{us}| f_+(0) \sqrt{I^\ell_K}$
in each decay process
is obtained using the experimental inputs
and correction factors in Refs.~\cite{Antonelli:2010yf,Seng:2021nar,Seng:2022wcw,Seng:2021boy,Seng:2021wcf}.
Figure~\ref{fig:vus_Ikl} shows that 
the six values of $|V_{us}|$ are consistent within the total error
evaluated using the lattice and experimental errors.
Their weighted average with the experimental error is also plotted.

\begin{figure}[!t]
 \centering
 \includegraphics*[scale=0.43]{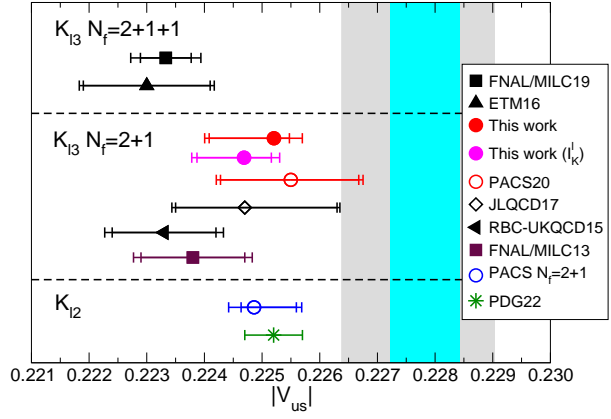}
 \caption{
Comparison of $|V_{us}|$ using our result of $I_K^\ell$ with 
previous results using $f_+(0)$~\cite{Boyle:2013gsa,Boyle:2015hfa,Aoki:2017spo,Carrasco:2016kpy,Bazavov:2018kjg,PACS:2019hxd,Ishikawa:2022otj}.
$|V_{us}|$ determined from the $K_{\ell 2}$ decay are also
plotted using $F_K/F_\pi$ in our calculation and 
PDG22~\cite{ParticleDataGroup:2022pth}.
The closed (open) symbols represent results 
in the continuum limit (at a finite lattice spacing).
The inner and outer errors
express the lattice QCD and total errors.
The total error is evaluated by adding the errors in the lattice QCD and experiment in quadrature.
The value of $|V_{us}|$ determined from the unitarity of the CKM matrix using $|V_{ud}|$ in Refs.~\cite{Seng:2018yzq} and \cite{Hardy:2020qwl} are presented by the light blue and gray bands, respectively.
  \label{fig:comp_vus}
 }
\end{figure}

The value of $|V_{us}|$ from the average of the six decay processes is 
compared with the previous results using $f_+(0)$~\cite{Boyle:2013gsa,Boyle:2015hfa,Aoki:2017spo,Carrasco:2016kpy,Bazavov:2018kjg,PACS:2019hxd} including
this calculation~\cite{Ishikawa:2022otj} as shown in Fig.~\ref{fig:comp_vus}.
The result using $I_K^\ell$ agrees with that using $f_+(0)$ in this work
within the total error.
It is also reasonably consistent with other previous results
and those determined from the kaon leptonic ($K_{\ell 2}$) decay 
using $F_K/F_\pi$ in our calculation and in 
PDG22~\cite{ParticleDataGroup:2022pth}.
On the other hand, our result with $I_K^\ell$ is 
4$\sigma$ (2$\sigma$) away from the value determined through
the CKM unitarity using $|V_{ud}|$ in Ref.~\cite{Seng:2018yzq} 
(\cite{Hardy:2020qwl}),
which is shown by the light blue (gray) band in the figure.

To clarify the difference,
it is important to reduce the uncertainty in our calculation.
As in our result of $f_+(0)$, the largest error in the $I_K^\ell$ result
stems from the fit form dependence in the continuum extrapolation.
This systematic error could be largely reduced, if data at a smaller lattice
spacing is added in the continuum extrapolation.
It is an important future work in our calculation.
To do this, we are generating the third PACS10 configuration at a smaller lattice spacing.

\section{Summary}

We have calculated the $K_{\ell 3}$ form factors in the $N_f = 2+1$
lattice QCD on large volumes at the physical point.
Using the form factors at the two lattice spacings, we have carried out
simultaneous $q^2$ interpolation and continuum extrapolation of the
form factors to obtain $q^2$ dependent form factors in the continuum limit.
From the $q^2$ dependence of the form factors, 
the slope and curvature at $q^2 = 0$ are evaluated.
Those results are consistent with the previous lattice calculations
and the experimental values.
The phase space integral is also evaluated using our form factors
in the continuum limit.
The value of $|V_{us}|$ determined using the results of 
the phase space integral is consistent with our result with $f_+(0)$
in the same calculation, and reasonably agrees with the previous results.
Since our error is dominated by the systematic error coming from
the fit form dependence in the continuum extrapolation, the reduction of
this error is an important future direction.
For this purpose, the third PACS10 configuration
is generated at a smaller lattice spacing.
We plan to calculate the $K_{\ell 3}$ form factors using the configuration.

\section*{Acknowledgments}
Numerical calculations in this work were performed on Oakforest-PACS
in Joint Center for Advanced High Performance Computing (JCAHPC)
under Multidisciplinary Cooperative Research Program of Center for Computational Sciences, University of Tsukuba.
This research also used computational resources of Oakforest-PACS
by Information Technology Center of the University of Tokyo,
and of Fugaku by RIKEN CCS
through the HPCI System Research Project (Project ID: hp170022, hp180051, hp180072, hp180126, hp190025, hp190081, hp200062, hp200167, hp210112, hp220079).
The calculation employed OpenQCD system\footnote{http://luscher.web.cern.ch/luscher/openQCD/}.
This work was supported in part by Grants-in-Aid 
for Scientific Research from the Ministry of Education, Culture, Sports, 
Science and Technology (Nos. 18K03638, 19H01892).
This work was supported by the JLDG constructed over the SINET5 of NII.

\bibliography{reference}

\end{document}